\def\year{2016}\relax
\documentclass[letterpaper]{article}
\usepackage{aaai16}
\usepackage{times}
\usepackage{helvet}
\usepackage{courier}
\usepackage[switch]{lineno}
\usepackage{algorithm}
\usepackage{algorithmic}
\usepackage{graphicx}
\usepackage{epstopdf}

\begin{document}
\title{Image Clustering without Ground Truth}
\author{Abhisek Dash$^1$, Sujoy Chatterjee$^2$, Tripti Prasad$^1$, \and Malay Bhattacharyya$^1$\\
$^1$Department of Information Technology\\
Indian Institute of Engineering Science and Technology, Shibpur, Howrah - 711103, India\\
E-mail: \{dash.ajayanu.abhisek, tripti.prasad5294\}@gmail.com, malaybhattacharyya@it.iiests.ac.in\\
$^2$Department of Computer Science \& Engineering\\
University of Kalyani, Nadia - 741235, India\\
E-mail: sujoy@klyuniv.ac.in
}
\maketitle

\begin{abstract}
Cluster analysis has become one of the most exercised research areas over the past few decades in computer science. As a consequence, numerous clustering algorithms have already been developed to find appropriate partitions of a set of objects. Given multiple such clustering solutions, it is a challenging task to obtain an ensemble of these solutions. This becomes more challenging when the ground truth about the number of clusters is unavailable. In this paper, we introduce a crowd-powered model to collect solutions of image clustering from the general crowd and pose it as a clustering ensemble problem with variable number of clusters. The varying number of clusters basically reflects the crowd workers' perspective toward a particular set of objects. We allow a set of crowd workers to independently cluster the images as per their perceptions. We address the problem by finding out centroid of the clusters using an appropriate distance measure and prioritize the likelihood of similarity of the individual cluster sets. The effectiveness of the proposed method is demonstrated by applying it on multiple artificial datasets obtained from crowd.
\end{abstract}

\section{Introduction}
Canvassing a large congregation of ideas, skills or participation often enhances the quality of content and idea generation. These mutual interactions and participation can be regarded as occurrences of collective intelligence. `Crowdsourcing' \cite{Howe:2006} is a well-known example of such collective intelligence. It leverages the wisdom of the pack and is already changing the way a mass of people produce knowledge, generate ideas and make them actionable. The most renowned example of crowdsourcing \cite{Ipeirotis:2010} is the distributed encyclopedia `Wikipedia'. Wikipedia, instead of creating an encyclopedia by hiring their own writers and editors, gives the authority to the crowd to create the information on their own. This crowd knowledge can often be directed towards some research problems in non-profit environments. In recent years, crowd intelligence has been effectively used in various research domains like image processing, natural language processing, sentiment analysis, etc.  The reason is that there are numerous existing real-life problems that cannot be solved by machines but incorporating the power of crowd can produce better solutions to us. Image annotations \cite{Whitehill:2009,NIPS2011_4187} is one of these types of task that becomes very hard for a computer to solve in a time efficient manner. But if the enormous human resources can be employed to cluster a large set of images to produce a clustering solution then the task might be completed very efficiently.

Clustering \cite{jain:99,Fred:2005} partitions the data objects based on the information that is reflected from the different data objects and their relationships. It is basically the grouping of similar type of objects depending upon some features. Again, the way of partitioning the objects into different groups is not same for different clustering algorithms. This causes to produce different clustering results even when the same dataset is given as input to different clustering algorithms. So, from this multiple different clustering solutions, it becomes very hard to predict the best one.

Clustering ensemble \cite{Strehl:2002,Fred:2005} is the traditional way of combining multiple clustering solutions to reach into a consensus decision. Over the years, a wide spectrum of clustering ensemble methods have been proposed to find the consensus from multiple clustering solutions. The optimal agreement is formulated as the partition that shares the most information with the ensemble of partitions, as measured by the Average Normalized Mutual Information (ANMI). In \cite{Strehl:2002} three heuristic consensus algorithms were introduced. These are based on graph partitioning, called Cluster-based Similarity Partitioning Algorithm (CSPA), Hyper Graph Partitioning Algorithm (HGPA), and Meta CLustering Algorithm (MCLA). But this clustering ensemble problem becomes highly challenging if the number of clusters becomes unknown as a prior knowledge and thus can be dissimilar in different clustering solutions.

There are very minimal study \cite{Ayad2008} is available in literature that is concerned with variable number of clusters and to find a consensus from those clustering solutions. The study by \cite{Ayad2008}, deals with the problem of variable number of clusters in clustering ensemble problem and they proposed three different methods based on cumulative voting to solve the problem.

In this paper, a crowd-powered clustering model is introduced to solve an image clustering task, which is challenging for a machine to solve, by finding a consensus solution from multiple crowd opinions. We have generated a few tricky images so that it can raise some kind of dilemma in the crowd workers while grouping similar kind of images into the same cluster. These images are posted without giving any prior knowledge about the expected (or true) number of clusters and the clustering solutions are solicited them. Therefore, each of the individual crowd worker can group the images from their individual perspectives about the features of the images. In this way, the solutions received from them contain variable number of clusters and the objective of the proposed approach is to find the consensus solution from multiple clustering solutions containing variable number of clusters. The effectiveness of the method has been demonstrated by utilizing the crowd-powered model to solve ambiguous image clustering problem that are hard for computers to solve.

\section{Preliminaries and Basic Definitions}
We introduce some basic terminologies that will be used throughout the paper. Since we are taking the response from different crowd workers which have nothing to do with other's responses, the number of cluster in each cluster sets can also be different. So, we need few indices which will depict the degree of similarity of these different cluster sets. We involve two terms, namely Rand Index \cite{rand1971} and Adjusted Rand Index \cite{hubert:1985}. These are defined below.

\textbf{Rand Index:} The \emph{Rand Index} \cite{rand1971} in data clustering is a measure of agreement between two data cluster sets. Given a set of $n$ objects $S = \{O_1, O_2, \ldots, O_n\}$ and two independent cluster sets of $S$ to be compared, namely $X = \{X_1, X_2, \ldots , X_p\}$ and $Y=\{Y_1, Y_2, \ldots, Y_q\}$, the Rand Index is computed as follows.

$$R(X,Y) = \frac{a+d}{a+b+c+d}.$$

Intuitively, $a+d$ can be considered as the number of agreements between $X$ and $Y$ and $b+c$ is the number of disagreements between them.

Here $a$ denotes number of pairs of objects in $S$ that are in same cluster in $X$ and in same cluster in $Y$. $b$ means number of pairs of objects in $S$ that are in same cluster in $X$ and in different clusters in $Y$. $c$ is the number of pairs of objects in $S$ that are in different clusters in $X$ and in same cluster in $Y$. Lastly, $d$ is number of pairs of objects in $S$ that are in different clusters in $X$ and in different clusters in $Y$.

\textbf{Adjusted Rand Index:} The \emph{Adjusted Rand Index} \cite{hubert:1985} measures the correspondence between two partitions on same data and removes the shortcomings of current version of the Rand Index. It uses a contingency table where each entry $n_{ij}$ denotes the number of objects which are present in cluster $X_i$ in cluster set $X$ and in cluster $Y_j$ in cluster set $Y$. Mathematically, $n_{ij} = |X_i \cap Y_j|$.

\section{Problem Formulation}
Let us formalize the clustering ensemble problem for variable number of clusters for a crowdsourced environment. Let $M = \{m_1, m_2, \ldots, m_p\}$ is a set of $p$ images and $Y$ be the set of $n$ crowd workers. Let $E = \{E_1, E_2, \ldots, E_n\}$ be the set of clustering solutions obtained from the set of $n$ crowd workers. Now, each of this $E_i$ denotes the partition of $m$ images into $k$ clusters such that $E_i = \{ m_1^{i}, m_2^{i}, \ldots, m_k^{i}\}$. Note that, each of this $E_i$ might contain different number of clusters $k$. Therefore, objective of the proposed approach is to find out the most optimal clustering solution such that the final solution is closest to all of the other solutions.

\section{Proposed Approach}
In this section, we first introduce a crowd-powered model to collect image clustering solutions with variable number of clusters from multiple ambiguous images, and then discuss the proposed method to find a consensus partition by combining these solutions.

\subsection{The Crowd-powered Model}
To develop a crowd-powered clustering model, a platform is initially designed for soliciting clustering opinions over some images from the crowd workers. To trap them into some dilemma about the grouping of the similar kind of images, some tricky images are posted there. Again, as no prior knowledge about the possible number of clusters is available to them, therefore, they have the freedom to choose any number of cluster and group them based on their individual perception. Just because the different crowd workers may choose different number of clusters, the clustering solutions obtained from them are of very much diverse in nature. So, the proposed method derives a robust consensus clustering solution from these diverse multiple clustering solutions.

In the crowd-powered model, we feed multiple questions, posted online for the crowd workers to respond, each containing a number of images. The crowd workers are required to suggest possible groups (clusters) with similar kind of images based on their individual perception. For the current analysis, three different questions are considered each containing five, seven and nine images, respectively. Basically, different crowd workers 
are asked to cluster the images depending on whatever criteria they feel to be appropriate. This induced different kinds of responses from the crowd workers, as we know different persons observe the same thing from different viewpoints. The crowd workers are instructed to label each image with an integer from 1 to the maximum number of clusters that they feel the set of images should belong to. The images which he/she thinks to be in the same cluster are given the same integer label.

For example, one of the given questions consists of seven images of different footballers each having a distinct role in the team, e.g., goalkeeper, striker or mid-fielder. The snapshot of a sample question is highlighted in Fig.~\ref{Fig:Image_clustering_data_sample}. So, in this question all the goalkeepers must be given the same label, and it also applies to the other group of players. But someone can also cluster the same set of images with a different perspective. Say, whether the particular player has won the prestigious FIFA Ballon d'Or award or not. Depending on the club or country for which they played, the list is endless. Thus, in this way the clustering solutions for the same set of images are obtained from different crowd workers. So, as different crowd workers can opt any number of clusters for grouping the images, therefore, it is expected that the different clustering solutions obtained from them might have different number of clusters.

\subsection{Proposed Consensus Model}
To make the consensus from multiple solutions, initially the clustering solutions are stored in a three dimensional jagged array (e.g., $CS[i][j][k]$). Where $CS[i]$ is the clustering solution obtained from the $i^{th}$ user, $CS[i][j]$ refers to the array of the objects in the response of the $i^{th}$ user which are clustered in the $j^{th}$ label and $CS[i][j][k]$ are the objects themselves. Here it is worth mentioning that, all the $CS[i][j]$s are the partitions of all the $m$ number of images. So, the intersection of any $CS[i][j]$ and $CS[i][k]$, where $j \ne k$, is always a null set. So, in this way the objects are basically partitioned into different sets of different cardinalities in the jagged array. The jagged array has nothing to do with the label correspondence except simplifying the implementation, as we need to compare the objects pairwise.

As there is no correspondence between any two clustering solutions, therefore, some label correspondence should be made to make the clustering solutions standardized. But the label correspondence is not needed in the initial part because of the fact that we are giving more priority to the pair correspondence, i.e., whether a given pair is in the same cluster or they are in different clusters. Let us assume the two clustering solutions \{1, 1, 1, 2, 2\} and another is \{2, 2, 2, 1, 1\}. Both the solutions are effectively same because they agree on the fact that object 1, object 2 and object 3 belong to the same cluster and object 4 and object 5 also belong to the same cluster. In this context, we have employed the Adjusted Rand Index (ARI) \cite{hubert:1985} as a similarity measure between a pair of clustering solutions. If the value of ARI is evaluated to be 1, then the clustering solutions are basically identical to each other. During the implementation of this part, the number of objects is taken as the only parameter without asking for the cluster labels.

After obtaining all the clustering solutions from the various crowd workers, the similarities among the solutions are found. So, if $n$ be the number of different clustering solutions for a question containing $p$ number of images, a similarity matrix of order $n \times n$ is principally generated. Now, using the ARI the pairwise similarity values are computed. Thus, for a particular clustering solution the pair-wise ARI value of other $n-1$ number of solutions are computed and summed up. Now, the clustering solution having the maximum aggregated value can be treated as a median clustering solution (centroid solution) as that means this clustering solution is more closed to the rest of the other clustering solutions. Again, as this median clustering solution has the maximum similarity value, therefore, the number of clusters in the median (centroid) solution can be treated as the near optimal number of cluster.

In order to map all the other clustering solutions in terms of the obtained centroid, a label correspondence matrix is designed for all the $n-1$ number of clustering solutions. In this way, the clusters can eventually be converted to fixed length clustering solutions, where the number of clusters becomes equal to the obtained centroid clustering solution. This can be done by using a probability matrix. By tracking how many cluster label $i$ is getting converted to label $j$ and finding the maximum probability in the $i^{th}$ row, the cluster label $i$ can be converted to the corresponding maximum index for the corresponding cluster set.

\section{Experimental Results}
We present experiments on three artificial datasets to evaluate the performance of the proposed algorithm (without having any ground truth about the number of clusters). Experiments are implemented in java version ``1.8.0-ea" Java(TM) SE Runtime Environment (build 1.8.0-ea-b94) Java HotSpot(TM) 64-Bit Server VM (build 25.0-b36, mixed mode) on an Intel core-i5 processor of Intel(R) Core™ i5-3337U CPU @ 1.80GHz.

\subsection{Dataset Details}
\begin{figure}[h]
\begin{center}
\includegraphics[height=2.3in,width=3.4in]{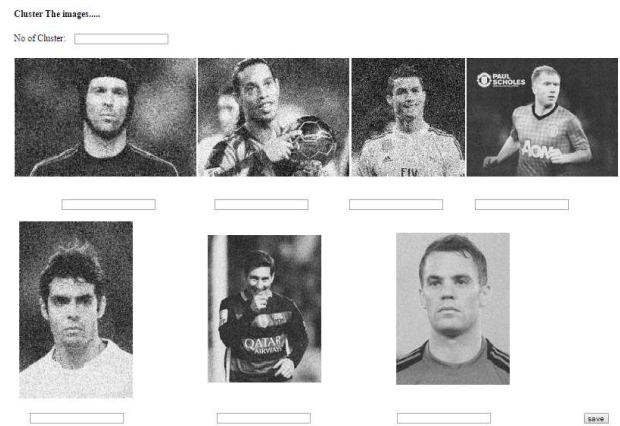}
\end{center}
\caption{Snapshot of a question for image clustering task through crowd opinions.}
\label{Fig:Image_clustering_data_sample}
\end{figure}

The data is generated from the crowd-powered platform discussed earlier. Consider a specific question where five images are shown in the webpage. The user is to label them using an integer value between 1 and the maximum number of clusters (they feel to be appropriate). Thus, the clustering solutions obtained from them may contain variable number of clusters. Suppose a crowd worker thinks that the images at the first, second and fourth position are in the same cluster, and the image at third position and fifth position are independently in different clusters. Then the data received will be \{1, 1, 2, 1 , 3\}. We used 3 such questions each containing several images to be clustered. A snapshot of the platform with the images is shown in Fig.~\ref{Fig:Image_clustering_data_sample}.

\subsection{Study on the Datasets}
As the different crowd workers registered their opinions (clustering solutions) for the images with varying number of clusters, therefore, their individual solutions can be compared with the consensus clustering solution produced by the proposed ensemble method. As rand index is used as the main performance metric, for a particular question if responses are collected from multiple crowd workers, then average rand index value can be computed. It will highlight how human computation can be effectively utilized for clustering these tricky and ambiguous images that are generally hard for the computer to cluster. Moreover, we can also compute the performance of the proposed approach for every question by comparing the clustering solution (derived by the proposed method) with respect to the ground truth clustering solution obtained from the question setter.

In order to investigate the efficiency of crowdsourced clustering containing solutions with variable number of partitions, we have computed the similarity of ensemble solution derived by applying the proposed method with the different clustering solutions obtained from crowd workers. In the Tables~\ref{tab:first_dataset_avg_result1}-\ref{tab:third_dataset_avg_result1}, these average adjusted rand index values are reported and a comparative performances with other state-of-the-art ensemble approaches are provided here. As most of the state-of-the-art ensemble approaches combine clustering solutions with fixed number of partitions, so we have applied those algorithms on the same input clustering solutions but with varying number of clusters (i.e., varying $k$). On the other hand, in this proposed ensemble method there is no need to supply the number of cluster as an input and it can be automatically generated therefore only one cell for proposed approach is filled up. It can be seen from the outputs of all datasets that the solution obtained from crowd workers provide equally competitive results without knowing the ground truth value of $k$. We treat the expert opinions (i.e., the opinions of question setter) as the ground truth label.

\begin{table}[!h]
\centering
\caption {Performance in terms of average adjusted rand index when the ensemble clustering solution is compared with each of the individual input solution (for $1^{st}$ dataset). 
} \label{tab:first_dataset_avg_result1}
 \begin{tabular}{|c | c | c | c | c  |}
 \hline
 Algorithm & $k=2$ & $k=3$ & $k=4$ & $k=5$ \\
 \hline
 CSPA & 0.1861 &  0.2170 &  0.3291 & 0.3291 \\
 \hline
 HGPA & 0.1861 &  0.3291 &  0.3291 & 0.5125  \\
 \hline
 MCLA & 0.1349 &  0.3291 &  0.5125 & 0.5125 \\
 \hline
 Proposed & --  & 0.3291 & -- & -- \\
\hline
\end{tabular}
\end{table}

\begin{table}[!h]
\centering
\caption {Performance in terms of average adjusted rand index when the ensemble clustering solution is compared with each of the individual input solution (for $2^{nd}$ dataset).} 
 \begin{tabular}{|c | c | c | c | c  |}
 \hline
 Algorithm & $k=2$ & $k=3$ & $k=4$ & $k=5$ \\
 \hline
 CSPA & 0.3821 &  0.6034 & 0.4441 & 0.5659 \\
 \hline
 HGPA & 0.3821 &  0.6034 & 0.4162 & 0.6034  \\
 \hline
 MCLA & 0.3821 &  0.6034 & 0.6034 & 0.4441 \\
 \hline
 Proposed & --  & 0.6034 & -- & -- \\
\hline
\end{tabular}
\end{table}

\begin{table}[!h]
\centering
\caption {Performance in terms of average adjusted rand index when the ensemble clustering solution is compared with each of the individual input solution (for $3^{rd}$ dataset). 
}
 \label{tab:third_dataset_avg_result1}
 \begin{tabular}{|c | c | c | c | c  |}
 \hline
 Algorithm & $k=2$ & $k=3$ & $k=4$ & $k=5$ \\
 \hline
 CSPA & 0.3103 &  0.5087 &  0.6909 & 0.5412 \\
 \hline
 HGPA & 0.3103 &  0.4680 &  0.6909 & 0.5412  \\
 \hline
 MCLA & 0.3103 &  0.5087 &  0.6909 & 0.6909 \\
 \hline
 Proposed & --  & -- & 0.6909 & -- \\
\hline
\end{tabular}
\end{table}

\section{Conclusions}
In this paper, our main focus is to utilize the power of crowd to perform image clustering without knowing the original number of clusters. We propose a method to combine multiple diverse clustering solutions (with variable number of clusters) to generate more robust solution. This problem is economically feasible as it can involve voluntary applications. For example, to search for exact location of the lost flight M370 of Malaysian Airlines, the volunteer based crowdsourcing was applied to group the wreckage of different portions of the fight to identify the location of the aircraft. Moreover, categorization of crowd workers based on their skill and diversity to filter out noise can be another future directions of research.
\section*{Acknowledgments}
The work of Malay Bhattacharyya is supported by the Visvesvaraya Young Faculty Research Fellowship 2015-16 of DeitY, Government of India. All the authors would like to thank the crowd contributors involved in this work.

\bibliographystyle{aaai}
\bibliography{clus_ensemble}

\end{document}